\documentclass[conference,a4paper]{IEEEtran}

\usepackage{cite}
\usepackage{amsmath}
\usepackage{amsxtra}
\usepackage{array}
\usepackage{mdwmath}
\usepackage{mdwtab}
\usepackage{eqparbox}
\usepackage{url}
\usepackage{graphicx}
\usepackage{subfig}
\usepackage{epsfig}
\usepackage{multirow}
\usepackage{textcomp}
\usepackage[keeplastbox]{flushend}

\begin{document}

\title{A Color Compensation Method Using Inverse Camera Response Function for Multi-exposure Image Fusion}

\author{\IEEEauthorblockN{Artit Visavakitcharoen, Yuma Kinoshita and Hitoshi Kiya}
\IEEEauthorblockA{Tokyo Metropolitan University, Hino, Tokyo 191-0065, Japan}
Email: visavakitcharoen-artit@ed.tmu.ac.jp, kinoshita-yuma@ed.tmu.ac.jp, kiya@tmu.ac.jp}
\maketitle


\begin{abstract}
Multi-exposure image fusion is a method for producing an image with a wide dynamic range by fusing multiple images taken under various exposure values. In this paper, we discuss color distortion included in fused images, and propose a novel color compensation method for multi-exposure image fusion. In the proposed method, an inverse camera response function (CRF) is estimated by using multi-exposure images, and then a high dynamic range (HDR) radiance map is recovered. The color information of the radiance map is applied to images fused by conventional multi-exposure imaging to correct the color distortion. The proposed method can be applied to any existing fusion approaches for improving the quality of the fused images.
\end{abstract}


\IEEEpeerreviewmaketitle

\section{Introduction}
The low dynamic range (LDR) of imaging sensors used in modern digital cameras is a major factor preventing cameras from capturing images as good as those with human vision. Accordingly, the interest of multi-exposure image fusion has recently been increasing. Various research works on multi-exposure image fusion have so far been reported \cite{mertens2007exposure,nejati2017fast,kinoshita2018automatic_trans,kinoshita2019scene}. These fusion methods utilize a set of differently exposed images, i.e. multi-exposure images, and fuse them to produce an image with high quality. However, conventional multi-exposure image fusion methods have not paid enough attention to the color of fused images, although they have paid attention to the spread of luminance.

Because of such a situation, we pointed out that multi-exposure images have different colors, so the fused images have to include some color distortion \cite{ClrCompensation2019IWAIT}. To improve this issue, we focus on two insights: the constant hue plane in RGB color space \cite{ueda2018contrastICIP} and inverse camera response function (CRF). In this paper, an inverse CRF is estimated by using multi-exposure images, and then a high dynamic range (HDR) radiance map is recovered to estimate the correct colors of a scene.  Next, the estimated color information is applied to a conventional  multi-exposure image fusion method on the constant hue plane in the RGB color space. The proposed method is not only a hue-preserving fusion method without gamut problem, but also a method applicable for any existing fusion methods to improve the quality of fused images.

\section{Proposed Method}
\label{sec:Proposed}

\subsection{Overview of Proposed Method}
\label{subsec:Overview}
\begin{figure}[!t]
  \centering
  \includegraphics[width=0.4\textwidth]{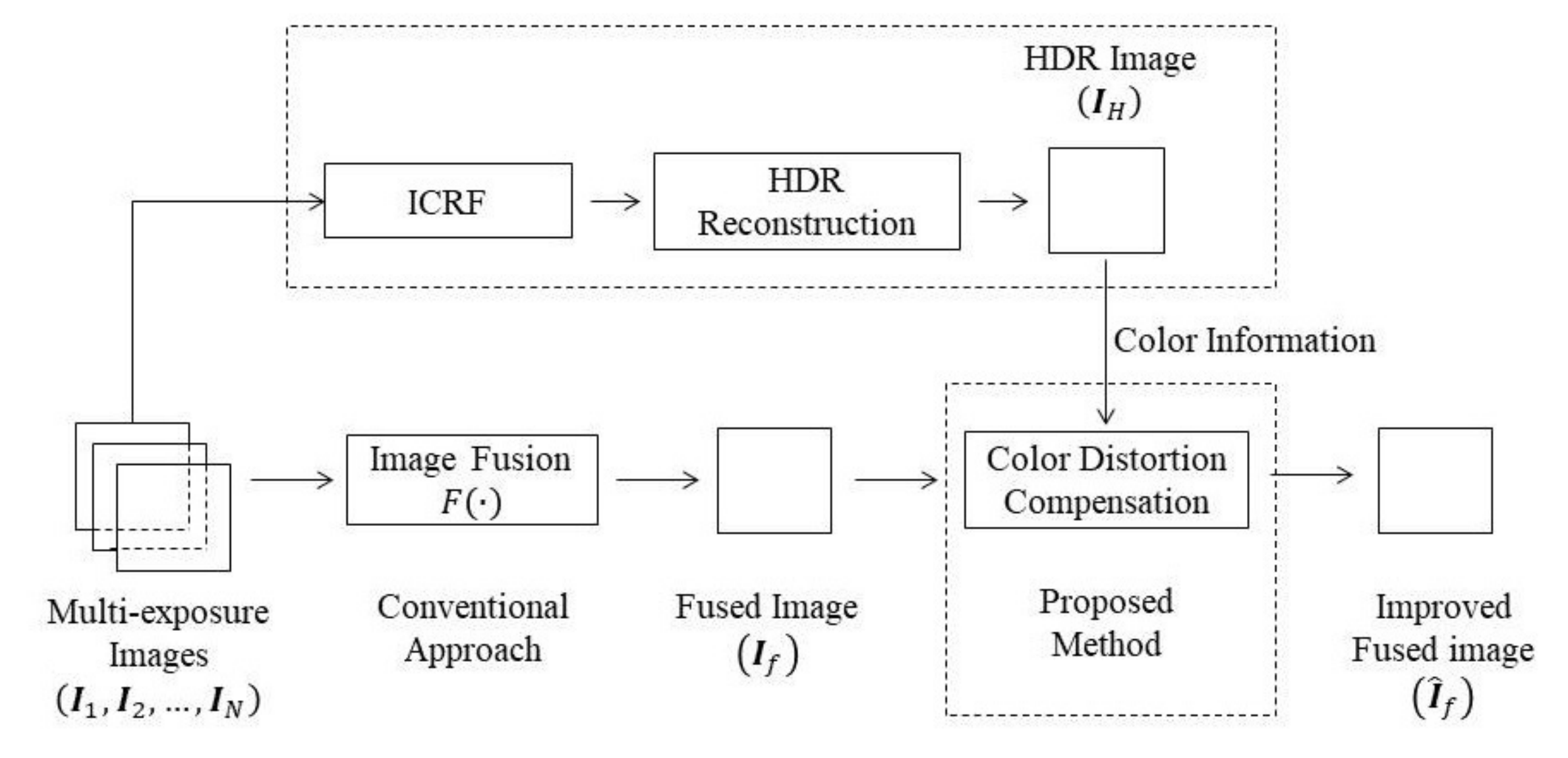}
  \caption{Proposed multi-exposure image fusion}
  \label{fig:BlockDiagram}
\end{figure}

The diagram of the proposed method is illustrated in Fig. \ref{fig:BlockDiagram}.

\vspace{-3mm}

To improve the quality of images fused by using a conventional image fusion method, we propose a color compensation method. Our approach is carried out as follows (See Fig. \ref{fig:BlockDiagram}).
\begin{enumerate}
	\item A fused image $\textbf{I}_f$ is produced by using a conventional image fusion method. It can be expressed as
	\begin{equation}
	\textbf{I}_f = f(\textbf{I}_1,\textbf{I}_2,...,\textbf{I}_N),
	\end{equation}
	where $f(\cdot)$ is an image fusion function and $\textbf{I}_1,\textbf{I}_2,...\textbf{I}_N$ are multi-exposure images.
	\item An HDR radiance map is recovered from the multi-exposure images by estimating the inverse camera response function (ICRF) \cite{ICRF_Debevec_97} and then utilizing the estimated ICRF to reconstruct an HDR image $\textbf{I}_H$.
	\item Color compensation is carried out to improve the color distortion of the fused image $\textbf{I}_f$. The constant hue plane in the RGB color space \cite{ueda2018contrastICIP} is used for computing the maximally saturated color $\textbf{c}_f$ from $\textbf{I}_f$, and $\textbf{c}_f$ is then replaced with a new one $\hat{\textbf{c}}_f$ calculated from $\textbf{I}_H$ to obtain an improved image $\textbf{I}'_f$.
\end{enumerate}

The following is the detail of the proposed color compensation i.e. 2) and 3).

\subsection{Inverse Camera Response Function}
\label{subsec:Proposed}
We focus on the relationships between real scene luminance and pixel values \cite{ICRF_Debevec_97}. Let a camera response function for mapping the scene radiance $E_{i}$ into a pixel value $x_{ij}$ be $g(\cdot)$. The pixel value $x_{ij}$ at a spatial index $i$ with an exposure index $j$ is written as
\begin{equation}
x_{ij} = g(E_i \Delta t_j),
\label{Eq:CRF}
\end{equation}
where $\Delta t_j$ is an exposure time for an exposure index $j$.

In order to reproduce an HDR image from the multi-exposure images, Eq. (\ref{Eq:CRF}) is solved to obtain the scene luminance map for each image by
\begin{equation}
\ln E_i = g^{-1} (x_{ij}) - \ln \Delta t_j,
\label{Eq:ICRF}
\end{equation}
where $g^{-1} (\cdot)$ is an inverse camera response function. The scene luminance maps are used to estimate an HDR image $\textbf{I}_H$ as
\begin{equation}
y_i = \frac{\sum^{N}_{j=1} \omega (x_{ij})  (g^{-1} (x_{ij}) - \ln \Delta t_j)}{\sum^{N}_{j=1} \omega (x_{ij})},
\label{Eq:HDR_rec}
\end{equation}
where $y_i$ is referred to as pixel intensity in $\textbf{I}_H$, and $\omega (x_{ij})$ is weight for each image. When the pixel value is closer to the middle of intensity range, which is given by $(x_{min}+x_{max})/2$, the higher weight is used \cite{ICRF_Debevec_97}.

\subsection{Constant Hue Plane in RGB Color Space }
\label{subsec:HuePlane}
\begin{figure}[!t]
  \centering
  \includegraphics[width=0.8\columnwidth]{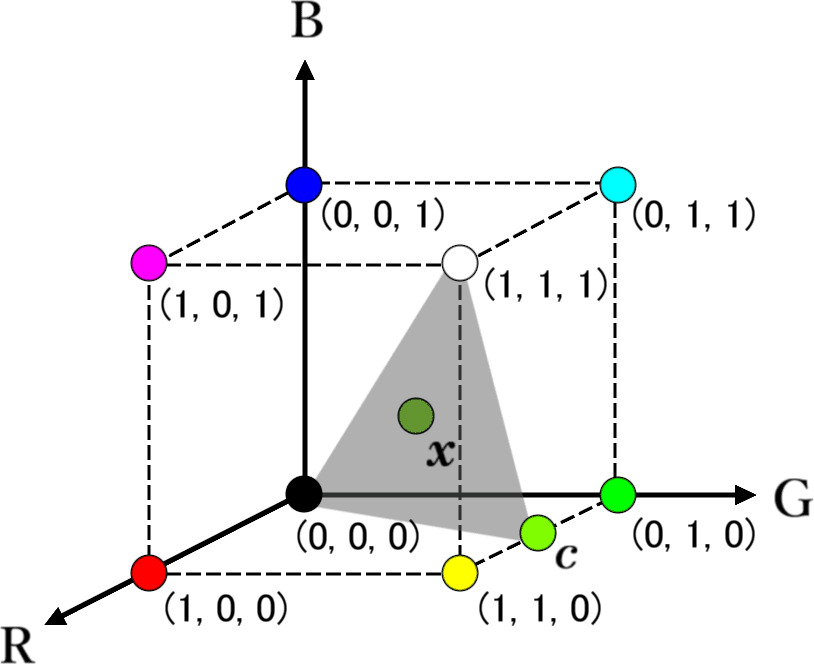}
  \caption{Constant hue plane in RGB color space}
  \label{fig:HuePlane}
\end{figure}

We utilize the constant hue plane in the RGB color space \cite{ueda2018contrastICIP}, as shown in Fig. \ref{fig:HuePlane}. Each pixel of an input image is represented in the RGB color space as $\textbf{x}=(a_r,a_g,a_b)$, $\textbf{x} \in [0,1]^3$. When a set of pixels has the same hue value, its intensity will align on the triangle, which consists of three vertices correspond to white, black and maximally saturated color represented by $\textbf{w}=(1,1,1)$, $\textbf{k}=(0,0,0)$ and $\textbf{c}=(c_r,c_g,c_b)$, respectively. The maximally saturated color $\textbf{c}$ is computed by
\begin{equation}
\label{Eq:c_all}
c_l =\frac{a_l-\min(\textbf{x})}{\max(\textbf{x})-\min(\textbf{x})}.
\end{equation}
where $l \in \{r,g,b\}$, and $\max(\cdot)$ and $\min(\cdot)$ are functions that return the maximum and minimum elements of the pixel $\textbf{x}$, respectively. Note that the elements of $\textbf{c}$ are in the range of $[0,1]$. When $\max(\textbf{x}) = \min(\textbf{x})$, i.e. $a_r=a_g=a_b$, the hue of the pixel $\textbf{x}$ is not defined.

The relationship between the RGB color space and the constant hue plane can be expressed by a linear combination with $\textbf{w}$, $\textbf{k}$, and $\textbf{c}$ components as
\begin{equation}
\begin{aligned}
\textbf{x} = a_r\textbf{r} + a_g\textbf{g} + a_b\textbf{b}, \\
= a_w\textbf{w} + a_k\textbf{k} + a_c\textbf{c}.
\label{Eq:Decomposition}
\end{aligned}
\end{equation}

Let pixels at the same location in $\textbf{I}_f$ and $\textbf{I}_H$ be $\textbf{x}_f$ and $\textbf{x}_H$, respectively, namely, as
\begin{equation}
\textbf{x}_f = a_{wf}\textbf{w} + a_{kf}\textbf{k} + a_{cf}\textbf{c}_f,
\label{Eq:Xf_Decomposition}
\end{equation}
\begin{equation}
\textbf{x}_H = a_{wH}\textbf{w} + a_{kH}\textbf{k} + a_{cH}\textbf{c}_H.
\label{Eq:XH_Decomposition}
\end{equation}
The proposed color compensation method is carried out by replacing $\textbf{c}_f$ with $\textbf{c}_H$, as
\begin{equation}
\hat{\textbf{x}}_f = a_{wf}\textbf{w} + a_{kf}\textbf{k} + a_{cf}\textbf{c}_H.
\label{Eq:Xfinal_Decomposition}
\end{equation}
By using this replacement, $\textbf{x}_f = a_{rf}\textbf{r} + a_{gf}\textbf{g} + a_{bf}\textbf{b}$ in the RGB color space is modified, as
\begin{equation}
\hat{\textbf{x}}_f = \hat{a}_{rf}\textbf{r} + \hat{a}_{gf}\textbf{g} + \hat{a}_{bf}\textbf{b}.
\label{Eq:Xfinal_RGB}
\end{equation}
The constant-hue plane have been studied for improving some color distortions
\cite{ClrCompensation2019IWAIT,kobayashi2019jpegxt}.

\section{Experiment}
\label{sec:Exp_Res}

In this experiment, 32 multi-exposure image sets were prepared form 32 HDR images $\textbf{I}_{H_0}$ \cite{HDRAny}. Each image set consists of 5 images with different exposure vales, i.e. $EV=[0,\pm0.5,\pm2]$ or $EV=[0,\pm1,\pm2]$.

In addition, we evaluated the color difference between $\textbf{I}_f$ and $\textbf{I}_{H_0}$ in terms of the difference of hue values $\Delta H_{ab}$ between two images based on the CIEDE 2000 color-difference formula \cite{CIEDE2004Sharma}, which was published by the CIE \cite{CIEDE_ISO}.

\begin{table}[!h]
	\centering
	\caption{Average values of $\Delta H_{ab}$ (with Mertens \cite{mertens2007exposure})}
	\label{tab:Mertens}
	\begin{tabular}{|c|c|c|}
		\hline
		EV Value & Method & $\Delta H_{ab}$ \\
		\hline
		\multirow{2}{*}{$EV=[0,\pm0.5,\pm2]$} & Conventional & 3.638 \\
		\cline{2-3}
		& Proposed with $\textbf{c}_H$ & \textbf{2.452} \\
		\hline
		\multirow{2}{*}{$EV=[0,\pm1,\pm2]$} & Conventional & 3.668 \\
		\cline{2-3}
		& Proposed with $\textbf{c}_H$ & \textbf{2.420} \\
		\hline
	\end{tabular}
\end{table}

As shown in Table \ref{tab:Mertens}, color distortions included in images generates by a conventional fusion method were improved by our proposed method.

\section{Conclusion}
\label{sec:Conclusion}
We proposed a color compensation method for multi-exposure image fusion using HDR images reconstructed by estimating inverse camera response functions. In an experiment, the proposed method was demonstrated to be effective in terms of the CIEDE 2000 color-difference formula.


\begin{thebibliography}{10}
\providecommand{\url}[1]{#1}
\csname url@samestyle\endcsname
\providecommand{\newblock}{\relax}
\providecommand{\bibinfo}[2]{#2}
\providecommand{\BIBentrySTDinterwordspacing}{\spaceskip=0pt\relax}
\providecommand{\BIBentryALTinterwordstretchfactor}{4}
\providecommand{\BIBentryALTinterwordspacing}{\spaceskip=\fontdimen2\font plus
\BIBentryALTinterwordstretchfactor\fontdimen3\font minus
  \fontdimen4\font\relax}
\providecommand{\BIBforeignlanguage}[2]{{%
\expandafter\ifx\csname l@#1\endcsname\relax
\typeout{** WARNING: IEEEtran.bst: No hyphenation pattern has been}%
\typeout{** loaded for the language `#1'. Using the pattern for}%
\typeout{** the default language instead.}%
\else
\language=\csname l@#1\endcsname
\fi
#2}}
\providecommand{\BIBdecl}{\relax}
\BIBdecl

\bibitem{mertens2007exposure}
T.~Mertens, J.~Kautz, and F.~V. Reeth, ``Exposure fusion,'' in
  \emph{Proceedings of the 15th Pacific Conference on Computer Graphics and
  Applications}, \hskip 1em plus 0.5em minus 0.4em\relax IEEE, 2007, pp.
  382--390.

\bibitem{nejati2017fast}
\BIBentryALTinterwordspacing
M.~Nejati, M.~Karimi, S.~R. Soroushmehr, N.~Karimi, S.~Samavi, and K.~Najarian,
  ``{Fast exposure fusion using exposedness function},'' in \emph{Proceedings
  of IEEE International Conference on Image Processing}, \hskip 1em plus 0.5em
  minus 0.4em\relax IEEE, Sep. 2017, pp. 2234--2238.
\BIBentrySTDinterwordspacing

\bibitem{kinoshita2018automatic_trans}
\BIBentryALTinterwordspacing
Y.~Kinoshita and H.~Kiya, ``{Automatic exposure compensation using an image
  segmentation method for single-image-based multi-exposure fusion},''
  \emph{APSIPA Transactions on Signal and Information Processing}, vol.~7, no.
  e22, Dec. 2018.
\BIBentrySTDinterwordspacing

\bibitem{kinoshita2019scene}
\BIBentryALTinterwordspacing
Y.~Kinoshita and H.~Kiya, ``{Scene Segmentation-Based Luminance Adjustment for Multi-Exposure
  Image Fusion},'' \emph{IEEE Transactions on Image Processing}, vol.~28,
  no.~8, pp. 4101--4116, Aug. 2019.
\BIBentrySTDinterwordspacing

\bibitem{ClrCompensation2019IWAIT}
A.~Visavakitcharoen, Y.~Kinoshita, and H.~Kiya, ``Pure-color preserving
  multi-exposure image fusion,'' in \emph{International Workshop on Advanced
  Image Technology (IWAIT) 2019}, vol. 11049, Jan. 2019, p. 110493X.

\bibitem{ueda2018contrastICIP}
Y.~Ueda, H.~Misawa, T.~Koga, N.~Suetake, and E.~Uchino, ``Hue-preserving color
  contrast enhancement method without gamut problem by using histogram
  specification,'' in \emph{2018 IEEE International Conference on Image
  Processing (ICIP)}.\hskip 1em plus 0.5em minus 0.4em\relax IEEE, Oct. 2018,
  pp. 1128--1127.

\bibitem{ICRF_Debevec_97}
P.~E. Debevec and J.~Malik, ``Recovering high dynamic range radiance maps from
  photographs,'' in \emph{Proceedings of the 24th annual conference on Computer
  graphics and interactive techniques (SIGGRAPH'97)}, Aug. 1997, pp. 369--378.

\bibitem{kobayashi2019jpegxt}
\BIBentryALTinterwordspacing
H.~Kobayashi and H.~Kiya, ``{JPEG XT Image Compression with Hue Compensation
  for Two-Layer HDR Coding},'' in \emph{Proceedings of IEEE International
  Conference on Consumer Electronics - Asia}, Bangkok, Jun. 2019. [Online].
  Available: \url{http://arxiv.org/abs/1904.11315}
\BIBentrySTDinterwordspacing

\bibitem{HDRAny}
``High dynamic range image examples,''
  \url{http://www.anyhere.com/gward/hdrenc/pages/originals.html}.

\bibitem{CIEDE2004Sharma}
G.~Sharma, W.~Wu, and E.~N. Dalal, ``The {CIEDE2000} color-difference formula:
  Implementation notes, supplementary test data, and mathematical
  observations,'' \emph{Color Research \& Application}, vol.~30, no.~1, pp.
  21--30, 2005.

\bibitem{CIEDE_ISO}
ISO/CIE,
``{ISO/CIE 11664-6:2014 Colorimetry-Part 6: CIEDE2000 Colour-Difference Formula},'' 2014.

\end{thebibliography}

\end{document}